\documentclass[twocolumn,trackchanges]{aastex701}

\usepackage{soul}

\begin{document}

\title{On the possibility of chemically driven convection in red giants. \\ Implications for the He-core flash and mixing above the Red Giant Branch Bump}

\author[orcid=0009-0000-3735-9744,sname='Ocampo']{M. Miguel Ocampo}
\affiliation{Instituto de Astrofísica de La Plata,CONICET-UNLP. Avenida Centenario (Paseo del Bosque) S/N, La Plata, B1900FWA, Buenos Aires, Argentina}
\affiliation{Facultad de Ciencias Astronómicas y Geofísicas, Universidad Nacional de La Plata, B1900FWA, Buenos Aires, Argentina}
\email[show]{mocampo@fcaglp.unlp.edu.ar}  

\author[orcid=0000-0001-8031-1957
,gname=Bosque, sname='Miller Bertolami']{Marcelo M. Miller Bertolami} 
\affiliation{Instituto de Astrofísica de La Plata,CONICET-UNLP. Avenida Centenario (Paseo del Bosque) S/N, La Plata, B1900FWA, Buenos Aires, Argentina}
\affiliation{Facultad de Ciencias Astronómicas y Geofísicas, Universidad Nacional de La Plata, B1900FWA, Buenos Aires, Argentina}
\email{mmiller@fcaglp.unlp.edu.ar}

\begin{abstract}
Turbulent mixing remains one of the primary uncertainties in the modeling of stellar interiors. In stellar evolution simulations, regions where mixing occurs are typically identified using instability criteria. A particularly interesting situation arises when nuclear reactions produce inversions in the mean molecular weight within stellar interiors. Under these conditions, the material can become unstable to either thermohaline  or a Rayleigh-Taylor instabilities. We demonstrate that the standard criterion adopted in stellar evolution calculations does not accurately distinguish between these two regimes.  We derive an alternative criterion and show that  chemically driven convection in stellar interiors might be viable under much smaller mean molecular weight inversions than it is normally assumed.  
We investigate whether inversions in the mean molecular weight can trigger chemically driven convection above the red giant branch bump (RGBB) or during the helium core flash.

We find that the inversion at the base of the convective envelope above the RGBB is too weak and short-lived to sustain steady-state convection. In contrast, rapid carbon production at the base of the He-flash–driven convective zone can maintain a steady chemically driven convective region. This process could significantly alter our understanding of the He-core flash and warrants further study. 
\end{abstract}

\keywords{convection, stars: interiors, stars: evolution, turbulence}


\section{Introduction}
\setcounter{footnote}{0}
The turbulent transport of heat and matter within stellar interiors is a critical yet challenging process to quantify. Its influence on temperature gradients and chemical stratification severely affects predictions for stellar structure and evolution \citep{2017LRCA....3....1K, 2020mdps.conf...69K, 2021FrASS...7...95X}. Modeling this transport is particularly difficult and has remained one of the key uncertainties in stellar physics since the realization that stars can have convectively unstable regions \citep{1906WisGo.195...41S, 1932Biermann, 1955ApJS....2....1H}. Direct application of hydrodynamical 2D and 3D numerical simulations to stellar evolution calculations is still computationally prohibitive. Consequently, stellar structure and evolution models continue to rely on computationally less demanding, semi-analytical approaches. Despite extensive efforts to study stellar convection \citep{1993ApJ...407..284G, 1996ApJ...473..550C, 2011A&A...528A..76C, 2015MNRAS.451.3354X, 2015ApJ...809...30A, 2022A&A...667A..96K}, the mixing length theory  \citep[MLT,][]{1925Prandtl,1932Biermann,2023Galax..11...75J},
 is still the standard choice in the field of stellar evolution. This is in no small part because, in spite of its simplicity, MLT shows a surprisingly good agreement with observations.

The treatment of convection in one-dimensional (1D) stellar evolution codes typically involves two main steps. First, stability criteria are applied to identify regions where convection, semiconvection, thermohaline mixing, or other forms of mixing, develop. Once these regions are identified, mixing velocities and thermal gradients are computed for each unstable layer using the MLT for convection, or appropriate recipes for semiconvection \citep{1983A&A...126..207L, 1985A&A...145..179L} and thermohaline mixing \citep{1980A&A....91..175K, 2007A&A...467L..15C,2011ApJ...728L..29T, 2013ApJ...768...34B}.

A particularly interesting situation arises when heavier materials are created or added on top of lighter ones\footnote{Technically, what matters is not the gradient of the mean molecular weight itself, but the so-called Ledoux term, $B$, as will be discussed in the next section.}. Under these conditions, layers that would be stable in the absence of chemical stratification can become unstable \citep{2004cgps.book.....W, Kippenhahn2013}. Such situations are not mere idealizations; they occur naturally in stars. Examples include the creation of C-rich material during the off-center helium-core flash in low-mass stars \citep{1967ZA.....67..420T, 1980A&A....91..175K}, or mixing immediately above the red giant branch bump \citep[RGBB, see][]{1985ApJ...299..674K, 2023ApJ...943...45M} due to the hydrogen created by the $^{3}\rm{He}(^3\rm{He}, 2p)^4\rm{He}$ reaction \citep{2006Sci...314.1580E, 2007A&A...467L..15C}. This situation can also develop in the cores of white dwarfs during crystallization \citep{1983A&A...122..212M, 1997ApJ...485..308I} or in the envelopes of white dwarfs and neutron stars due to solid accretion from debris disks \citep{2013A&A...557L..12D, 2017A&A...601A..13W, 2023ApJ...950...73F}.

Mixing in most of these cases is typically attributed to thermohaline processes. Thermohaline instability differs from convective instability in that it requires two components diffusing at different speeds. In stars, the stabilizing component (heat/temperature) diffuses faster than the other (chemical composition). Under this condition, a fluid element displaced from its equilibrium position, which would be stable under adiabatic perturbations, becomes heavier than its surroundings as it loses heat via radiation and conduction. The result is that fluid elements sink or rise on a thermal timescale, which is much longer than the convective turnover timescales ruled by the local free-fall timescale. \cite{2024ApJ...969...10C} have shown that the simple expressions usually adopted in stellar evolution computations to describe both thermohaline mixing \citep{1980A&A....91..175K} and convection \citep{Kippenhahn2013}  can be obtained from the same set of equations.
The main difference between both sets of solutions is that thermohaline mixing is characterized by slow movements and almost no convective transport of energy, and proper convection in deep stellar interiors is characterized fast movements and temperature gradients close to adiabatic. Moreover, while typical mixing lengths in the case of stellar convection are of the order of the local pressure scale height, mixing lengths in thermohaline processes are many orders of magnitude smaller \citep{2020mdps.conf...13G}. 

While mixing in stars above the RGBB and in the layers below the off-centered He core flash is usually  discussed in the context of thermohaline mixing, the reality is far more complicated, with the possibility of convective elements with sizes of the order of the pressure scale height being involved. Notably, 2D and 3D hydrodynamical simulations of both the helium-core flash \citep{2011ApJ...743...55M} and the RGBB \citep{2006Sci...314.1580E} show  large convective elements with much faster motions, closer to those expected for Rayleigh-Taylor instabilities. Moreover, for the RGBB, the mixing efficiency of thermohaline processes required to explain the observed abundances is much higher than that expected from hydrodynamical simulations of the process, casting serious doubts on the nature of the actual mixing process \citep{2007A&A...467L..15C, 2010ApJ...723..563D, 2011ApJ...727L...8D, 2022ApJ...941..164F}. Recent works, however, seem to close this gap, showing that vertical magnetic fields enhance thermohaline mixing in 3D simulations \citep{2019ApJ...870L...5H,2024ApJ...964..184F} enough to match observations.

In this work, we study the predictions of combining the Ledoux criterion with an extension of MLT that includes the impact of chemical gradients on the development of convection in stellar interiors. We find that fast and almost adiabatic chemically driven convection is viable in the presence of much smaller chemical gradients than those indicated by the instability criterion usually adopted in stellar evolution codes.
Above a threshold chemical gradient, a much faster form of mixing is possible, which is similar to that of standard convective motions.
This finding suggests that the fast mixing processes observed in hydrodynamical simulations below the He-core flash convective zone \citep{2011ApJ...743...55M} and at the RGBB \citep{2006Sci...314.1580E} could be caused by chemical gradient crossing this critical threshold. We perform experiments to test if such scenarios could be modeled by a steady convective model like the MLT.

\section{Instability Criteria in 1D Stellar Evolution Codes}
To identify regions where mixing occurs, 1D stellar evolution codes rely on instability criteria. In the presence of chemical gradients, the relevant instability criterion is the Ledoux instability criterion \citep{Kippenhahn2013, 2004cgps.book.....W}, given by
\begin{equation}
  \nabla > \nabla_{\rm ad} + B.
  \label{eq:Ledoux_criterion}
\end{equation}
This criterion identifies the conditions under which a fluid element, adiabatically displaced from its equilibrium position, would be accelerated away by buoyancy forces. Here, $\nabla = \mathrm{d\ln}T/\mathrm{d\ln}P$ is the actual temperature stratification of the star, and $\nabla_{\rm ad} = \mathrm{d\ln}T/\mathrm{d\ln}P|_{\rm ad}$ is the adiabatic temperature gradient. In Eq. \ref{eq:Ledoux_criterion}, $B$ is the Ledoux term, which is defined as \citep{1989nos..book.....U}
\begin{equation}
  B = \frac{1}{\chi_T}\sum^{N-1}_{i=1} \chi_{X_i} \frac{\mathrm{d\ln}X_i}{\mathrm{d\ln}P},
  \label{eq:Ledoux_term}
\end{equation}
where $\chi_T=\partial\mathrm{\ln}P/\partial\mathrm{\ln}T|_{T,\{X_i\}}$ and $\chi_{X_i}=\partial\mathrm{\ln}P/\partial\mathrm{\ln}X_i|_{ T,\{X_j\}\neq i}$ are the thermal and generalized compressibilities, and $N$ is the number of chemical species in the mixture. Eq. \ref{eq:Ledoux_criterion} reduces to the classical Schwarzschild criterion \citep{1906WisGo.195...41S} when $B=0$, e.g. in a fully ionized homogeneous medium.

The application of Eq. \ref{eq:Ledoux_criterion} requires prior knowledge of the temperature gradient, $\nabla$, which, in turn, depends on whether turbulent heat transport is occurring in that layer. Indeed,  in stellar evolution computations, $\nabla$ is one of the main variables one wants to compute with a theory of convection. For this reason, convective zones are typically identified not by Eq. \ref{eq:Ledoux_criterion}, but by checking if the temperature gradient required to transport all heat by radiation or conduction ($\nabla_{\rm rad}$) would be unstable against adiabatic perturbations, i.e.
\begin{equation}
  \nabla_{\rm rad} > \nabla_{\rm ad} + B.
  \label{eq:False_Ledoux_criterion}
\end{equation}
It has been a common assumption in textbooks \citep{Kippenhahn2013, 2004cgps.book.....W, 2023Galax..11...75J, 2005essp.book.....S} that Eq. \ref{eq:False_Ledoux_criterion} provides a good approximation to the actual Ledoux criterion, and, in most cases, no distinction is made between the two. Indeed, to our knowledge, all state-of-the-art 1D stellar evolution codes that rely on MLT and consider the impact of chemical gradients use Eq. \ref{eq:False_Ledoux_criterion} \citep{2008Ap&SS.316...99W, 2014A&A...569A..63G, 2018ApJS..234...34P, 2022ApJ...928L..10A, 2023A&A...680A.101S}.

As we will discuss in the next section, we have found that this seemingly innocuous assumption  may have critical consequences. We have found that, for given values of $\nabla_{\rm ad}$, $\nabla_{\rm rad}$,  chemically driven convection might be viable at much lower values of $|B|$ than  indicated by the Eq. \ref{eq:False_Ledoux_criterion}. 

\section{Steady State Mixing in the Presence of Chemical Gradients}
The standard treatment of MLT found in textbooks \citep{Kippenhahn2013, 2004cgps.book.....W, 2023Galax..11...75J} can be extended to deal with convection in the presence of chemical gradients ($B \neq 0$). Under these conditions, it is straightforward to derive that the temperature changes of a moving fluid element ($\nabla_{\rm e} = \mathrm{d\ln}T/\mathrm{d\ln}P|_{\rm e}$) and the final temperature gradient of the background matter ($\nabla$) are related to $\nabla_{\rm ad}$, and $\nabla_{\rm rad}$ by the following equations \citep{1988Ap&SS.150..115U, 2024ApJ...969...10C}:
\begin{equation}
  \big(\nabla_{\rm e}-\nabla_{\rm ad}\big)\big(\nabla-\nabla_{\rm e}-B\big)^{1/2}=2U\big(\nabla-\nabla_{\rm e}\big)
  \label{MLT1}
\end{equation}
and
\begin{equation}
  \big(\nabla-\nabla_{\rm e}-B\big)^{1/2}(\nabla-\nabla_{\rm e})=\frac{8}{9}U\big(\nabla_{\rm rad}-\nabla) ,
  \label{MLT2}
\end{equation}
 and the mixing velocity of the moving elements is given by
\begin{equation}
    v^2 = g\frac{\chi_T}{\chi_\rho} \bigg(\nabla - \nabla_{\rm e}-B\bigg)\frac{l_m^2}{8H_P}
    \label{eq:velocity}
\end{equation}
Here, $g$ is the local gravity, $l_m$ is the mixing length, $H_P$ is the pressure scale height, and $\chi_\rho=\partial\mathrm{\ln}P/\partial\mathrm{\ln}\rho|_{T,\{X_i\}}$. The dimensionless quantity $U$ is defined as
\begin{equation}
    U\equiv \bar{U}\left(\frac{H_P}{l_m}\right)^2\equiv\frac{3acT^3}{c_P\rho^2 \kappa l_m^2}\sqrt{\frac{8H_P}{g\delta}},
    \label{eq:U_def}
\end{equation}
where $a$ is the radiation constant, $c$ is the speed of light, $c_P$ is the specific heat at constant pressure, $\rho$ is the density, and $\kappa$ is the Rosseland opacity.  Note that we have also defined the auxiliary quantity $\bar{U}$ which is independent from $l_m$. Under the standard assumption that the size of the convective elements is of the order of their mixing length, the value of $U$ is on the order of the ratio between the free-fall and thermal-adjustment timescales \citep{Kippenhahn2013}. Thus, $U$ and $\bar{U}$ are quantities that indicate how fast the convective element will thermalize to its surrounding. For smaller values of $U$, typical for dense media, radiative losses are not important compared to the convective motions. Greater values of $U$, on the contrary, indicate that the element, regardless of its velocity, will lose practically all of their excess heat via radiation. This is the case in regions of small density. However, $U$ also depends on the mixing length (and thus the size of the moving element). This could mean that, even in very dense matter ($\bar{U} \ll 1$), a small ``blob'' will radiate all of its excess heat. In stellar interiors, typically $\bar{U}\ll 1$. $U$ (and $\bar{U}$) are quantities also related with the convective efficiency $\Gamma=(\nabla-\nabla_{\rm e})/(\nabla_{\rm e}-\nabla_{\rm ad})$, used often as the variable to solve in MLT equations \citep{1988Ap&SS.150..115U, 2024ApJ...969...10C},   by $\Gamma=(\nabla-\nabla_{\rm e}-B)^{1/2}/2U$.

 In convective regions, the typical mixing length is of the order of the pressure scale height \citep[$l_{\rm conv}=\alpha \, H_P$, where $\alpha$ is a dimensionless number of order 1, see][]{Kippenhahn2013}.  Noteworthy, this situation is very different for thermohaline instabilities, where the typical lengths are orders of magnitude smaller than $H_P$. In this regime, typical length are given by $l_{\rm therm}= (\kappa_T\,\nu/{N_T}^2)^{1/4}$ 
 where $\kappa_T$ is the thermal diffusivity, $\nu$ the kinematic viscosity, and ${N_T}^2$ the squared thermal Brunt-Väisälä frequency \citep{2018AnRFM..50..275G, 2020mdps.conf...13G}.

At each layer of the star, for given values of  $W=\nabla_{\rm rad}-\nabla_{\rm ad}$, $B$, and $U$, Eqs. \ref{MLT1} and \ref{MLT2} allow us to find the resulting mean temperature gradients of the star ($\nabla$) and the moving element ($\nabla_{\rm e}$). Note that Eqs. \ref{MLT1} and \ref{MLT2}, (as well as Eq. \ref{eq:velocity}) have real solutions when
\begin{equation}
    \nu^2 = \nabla - \nabla_{\rm e} - B > 0,
    \label{eq:real_instability}
\end{equation}
which describes the situation where the fluid element is unstable against non-adiabatic perturbations.

It is important to note that Eqs. \ref{MLT1} and \ref{MLT2} do not always have a unique solution \citep{1988Ap&SS.150..115U}, meaning that in some regions of the $W\hbox{--}B$ parameter space, different mixing regimes can coexist. In such circumstances, the state of the fluid will depend on the initial conditions. When the Ledoux term $B<0$, there is always at least one real solution, indicating that some mixing always occurs within this picture. Note, however, that detailed studies of thermohaline mixing show that there is indeed a critical value of $|B|$  below which such mixing is not possible \citep{2013ApJ...768...34B,2011ApJ...728L..29T}. This limit is defined by the chemical diffusion, and cannot be captured by our simple model in which the displaced elements maintain their chemical composition.

\subsection{Solving the extended MLT equations}

\begin{figure}
   \centering
    \includegraphics[width=\columnwidth]{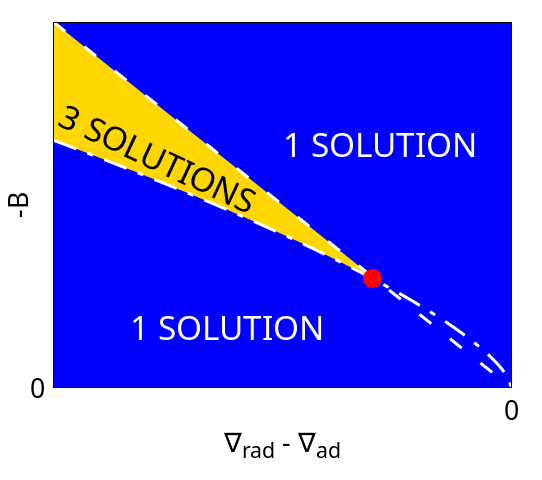}
    \caption{Schematic representation of the different regions in the $(\nabla_{\rm rad}-\nabla_{\rm ad} )-B$ space where eqs.  \ref{MLT1} and \ref{MLT2} have one or three solutions. In the figure only the quadrant where $W=(\nabla_{\rm rad}-\nabla_{\rm ad})<0$ and $B<0$ is displayed.}
    \label{fig:scheme}
\end{figure}

The resolution of Eqs. \ref{MLT1} and \ref{MLT2} in the conditions of stellar interiors $U\ll 1$ can be numerically challenging without a proper rewritting of the equations. This is a consequence of the fact that the temperature gradients $\nabla$ and $\nabla_{\rm e}$, that one wants to finds for given values of  $\nabla_{\rm rad}$, $\nabla_{\rm ad}$ and $U$, can converge to almost adiabatic ($\nabla\simeq \nabla_{\rm e}\simeq\nabla_{\rm ad}$) or almost radiative solutions ($\nabla\simeq \nabla_{\rm e}\simeq\nabla_{\rm rad}$) to extreme levels of precision. Consequently, neither  $\nabla$ or $\nabla_{\rm e}$ are good variables to numerically solve these equations. A better option is the dimensionless velocity $\nu$.  Eqs. \ref{MLT1} and \ref{MLT2} can be combined into a single quartic equation for  $\nu$ given by
\begin{eqnarray} \label{eq:velocityQ}
Q(\nu) &=& \nu^4+\frac{8U}{9}\nu^3+\left(B+\frac{16U^2}{9}\right)\nu^2 \nonumber \\ &+&\frac{8U}{9}(B-W)\nu
+\frac{16 U^2}{9}B=0.
\end{eqnarray}  
Noteworthy the (positive) roots of Eq. \ref{eq:velocityQ} can differ by many orders of magnitude. For this reason, in order to solve Eq. \ref{eq:velocityQ} a change of variables $\nu=\exp{x}$ is convenient. This change of variables also removes from the treatment the negative solutions for $\nu$, which are unphysical. Once this is done, the roots can be easily found by standard numerical methods.

In our region of interest in the $W-B$ space, that is when $W<0$ (stable according to Schwarzschild criterion) and $B<0$, Eq. \ref{eq:velocityQ} has either one or three real positive solutions, depending on the values of $W$, $B$, and $U$. Fig. \ref{fig:scheme} shows schematically where the different regimes are located. At large values of $|W|$, i.e. to the left of the red point in Fig. \ref{fig:scheme} the region where three different solutions exist is delimited by two curves, which we will discuss later (see Section \ref{sub:critical}). For $U\ll 1$, a condition typical of stellar interiors when $l_m \not \ll H_P$ ,  the region to the right of the red point in Fig. \ref{fig:scheme} is confined to extremely low values of $W$, which makes the region completely irrelevant, as in that case $\nabla_{\rm rad}\simeq \nabla_{\rm ad}$.

\begin{figure*}
     \centering
    \includegraphics[width=\textwidth]{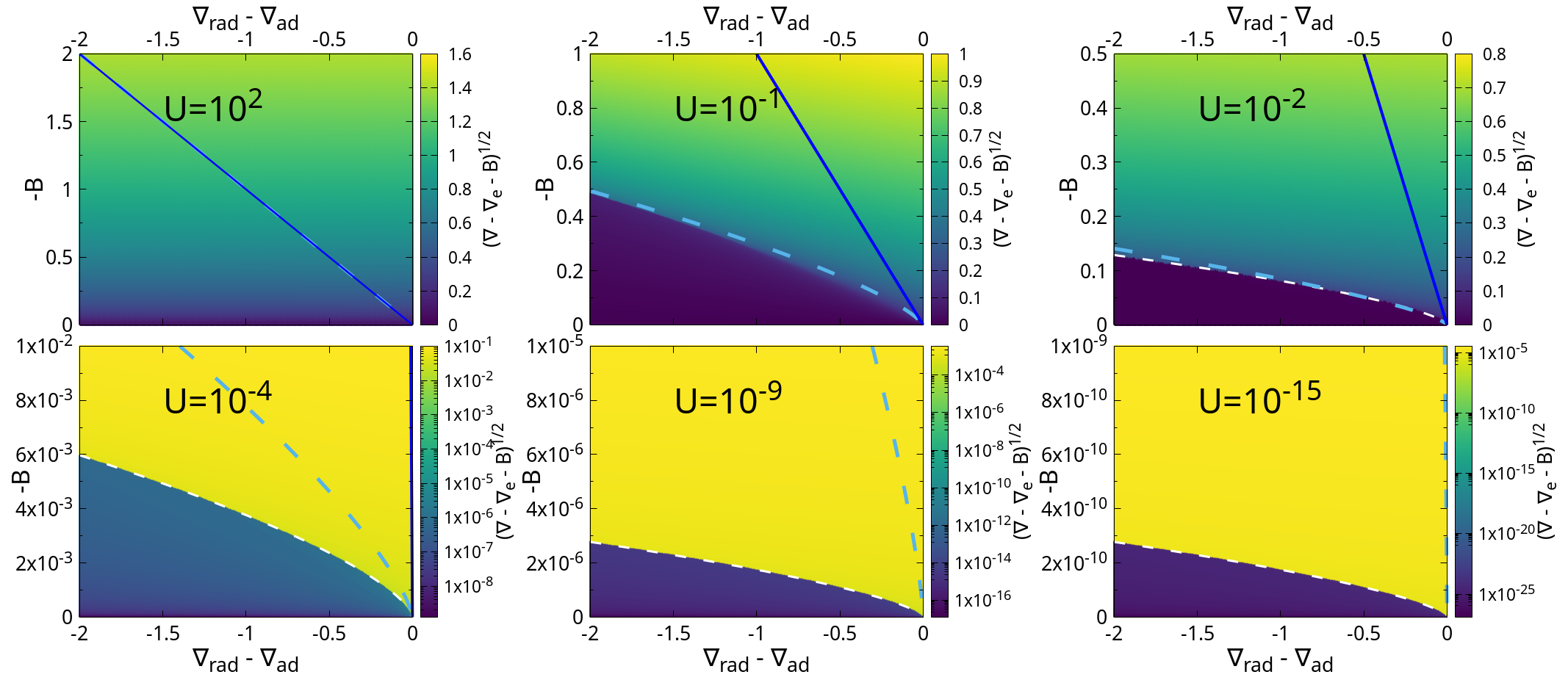}
    \caption{Values of the dimensionless velocity $\nu=(\nabla - \nabla_{\rm e}-B)^{1/2}$ for different values of $U$. The dashed cyan lines indicate the location where the intermediate solution (or the only solution for large values of $U$) fulfills $\nabla=\nabla_{\rm ad}+B$ (neutrality according to the  Ledoux criterion). The criterion usually adopted in stellar evolution computations, given by Eq. \ref{eq:False_Ledoux_criterion}, is shown as a solid blue line. Dashed white lines in the bottom three and top right panels ($U\ll1$) indicate the location of Eq. \ref{eq:Real_Criterion}, which perfectly reproduces the location of the critical line dividing thermohaline-like and convective solutions, while the standard criterion given by Eq. \ref{eq:False_Ledoux_criterion} basically overlaps with the vertical axis at that scale. 
    }
    \label{fig:velocities}
\end{figure*}

When three solutions of Eq. \ref{eq:velocityQ} for $\nu$ coexist, they can be classified as slow, intermediate and fast velocity solutions. The coexistence of, at least, two types of mixing in the presence of inverse chemical gradients was previously noted by \cite{2010ApJ...723..563D}. Also, the slow solutions will always be stable against the actual Ledoux criterion, and the fast will always be unstable. The intermediate solution, on the other hand, is the only one that transitions from stable to unstable and, then, satisfies the neutral Ledoux condition at some point. \cite{2024ApJ...969...10C} have shown that the low-velocity solutions correspond to thermohaline mixing and match the velocities originally derived by \cite{1980A&A....91..175K} when $|B|\ll|W|$.

Fig. \ref{fig:velocities} shows the predictions of Eqs. \ref{MLT1} and \ref{MLT2} for the dimensionless mixing velocity $\nu$ (or the largest solution when multiple are present) for different values of $W$, $B$, and $U$. Several key features are apparent from this figure. First, for low values of $U$ there is a sharp separation between the slow and fast velocity regimes. Second, it is clear that the criterion from Eq. \ref{eq:False_Ledoux_criterion} does not identify such transition, and the discrepancy becomes more evident with lower values of $U$. Third, we can see that, for higher values of $U$, there is no transition from slow to fast regime, since for the values considered for $W$ there is only one solution to the MLT equations. The actual Ledoux criterion is also displayed, and can be noted that for higher values of $U$ it merges with Eq. \ref{eq:False_Ledoux_criterion}, since $\nabla \rightarrow \nabla_{\rm rad}$ as $U\rightarrow\infty$ \citep{Kippenhahn2013}. The Ledoux criterion is obtained from Eqs. \ref{MLT1} and \ref{MLT2} in Appendix \ref{methods}.

\begin{figure*}
    \centering
    \includegraphics[width=\textwidth]{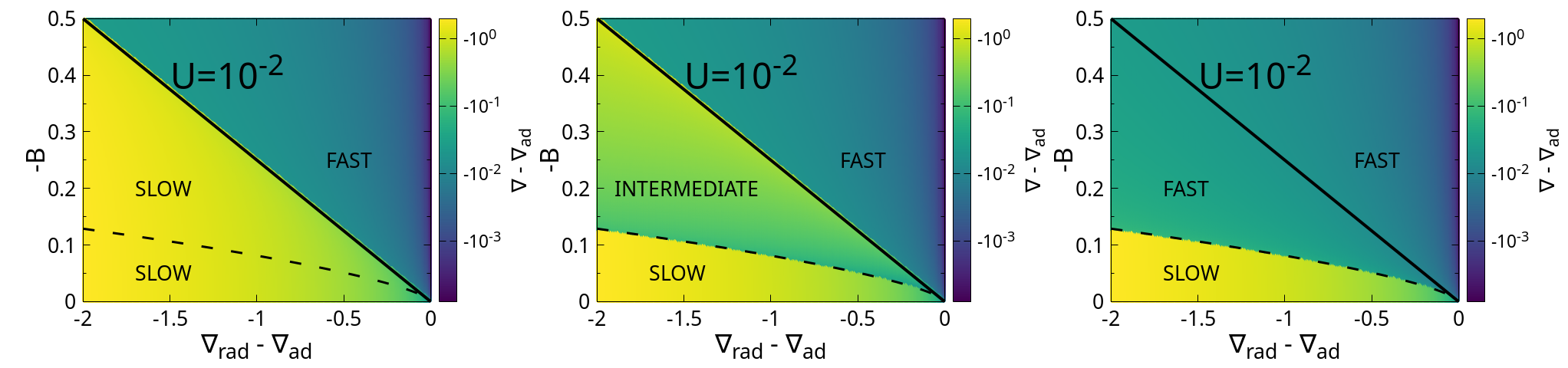}
    \includegraphics[width=\textwidth]{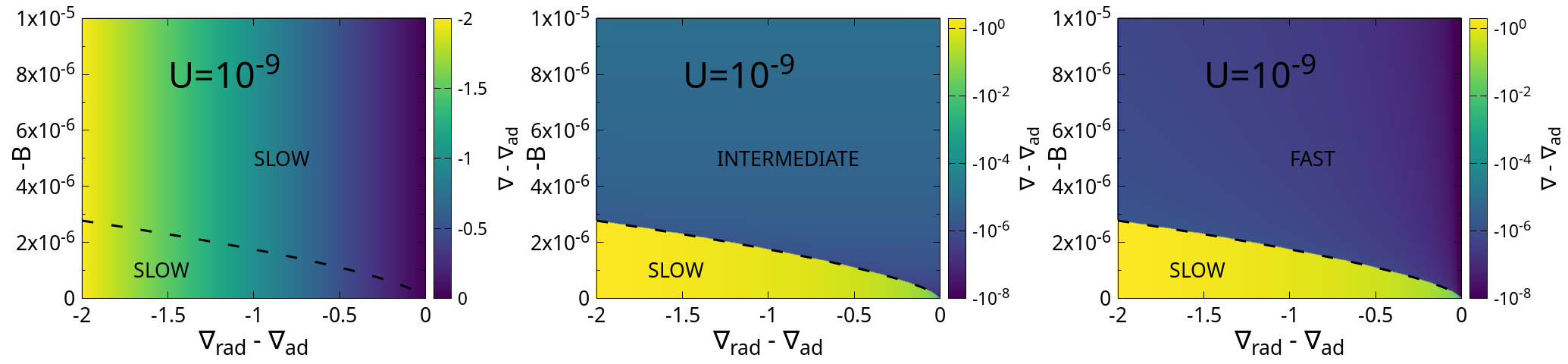}
    \caption{Values for the sub-adiabacity $\nabla-\nabla_{\text{ad}}$ for $U=10^{-2}$ and $U=10^{-9}$ when the slow (left panels), intermediate (middle panels) or fast solution (right panels) is chosen in the region between Eqs. \ref{eq:uppercurve} (solid) and \ref{eq:Real_Criterion} (dashed). In the bottom panels, Eq. \ref{eq:uppercurve} is not displayed since it overlaps with the vertical axis at that scale. Note that the color bar range of the lower-left panel is linear to highlight the radiative behavior of the solutions ($\nabla-\nabla_{\text{ad}}\simeq \nabla_{\text{rad}}-\nabla_{\text{ad}}$) in the slow velocity regime.}
    \label{Fig:SAD}
\end{figure*}

 In our region of interest, $\{\nabla_{\text{rad}}<\nabla_{\text{ad}},B<0\}$, all the solutions will be sub-adiabatic ($\nabla<\nabla_{\text{ad}})$. In contrast to Schwarzschild unstable layers, the gradients under these conditions satisfy $\nabla_{\text{rad}}<\nabla<\nabla_{\text{e}}<\nabla_{\text{ad}}$. Consequently, the convective heat flux, given by $F_c=\rho vc_P T(\nabla-\nabla_{\text{e}})l_m/H_P$ will always be negative and therefore inward. However, the magnitude of the sub-adiabaticity $\nabla-\nabla_{\text{ad}}$ will depend on the type of solution considered (slow, intermediate, or fast) and the value of $U$. Fig. \ref{Fig:SAD} shows $\nabla - \nabla_{\text{ad}}$ for $U=10^{-2}$ and $U=10^{-9}$ considering all the possible solutions in the region. 
 As we will discuss in Section \ref{sub:critical} the regions are defined by two critical lines, the dashed and continuous line in Fig. \ref{Fig:SAD}  (see also Fig. \ref{fig:scheme}). In between these two lines, three solutions coexist (namely slow, intermediate and fast solutions), whereas outside this region only one solution exists. Fig. \ref{Fig:SAD} shows the value of the sub-adiabaticity when either the slow, intermediate or fast solutions are chosen for the in-between region. Choosing the slow solution shows values of $\nabla-\nabla_{\text{ad}}$ continuous with the values below the dashed line, where only slow solutions exist, whereas choosing the fast one shows continuity with the gradients above the continuous line, where only fast solutions exist. The intermediate solution is not continuous with either slow nor fast solution gradients.
 
 As we stated at the beginning of this subsection, we can see from Fig. \ref{Fig:SAD} that $\nabla$ approaches $\nabla\simeq\nabla_{\text{rad}}$ for the slow solutions, approaches $\nabla\simeq\nabla_{\rm ad}$  for the fast solutions, and takes intermediate values for the intermediate solutions. The radiative temperature stratification of the slow solutions can be graphically appreciated especially in the bottom left panel of Fig. \ref{Fig:SAD}, where we did not use a logarithmic scale on the color map, and it can be clearly seen that $\nabla-\nabla_{\text{ad}}\simeq\nabla_{\text{rad}}-\nabla_{\text{ad}}$. The slow velocity of the solutions means that $\nu^2=(\nabla-\nabla_{\text{e}}-B)\simeq0$, which implies that $\nabla-\nabla_{\text{e}}\simeq B$. Additionally, in the $U\ll1$ regime, the region with only slow solutions correspond to $|B|\ll 1$ (see bottom panels of Fig. \ref{Fig:SAD}), which translates into $\nabla\simeq\nabla_{\text{e}}\simeq\nabla_{\text{rad}}$. We emphasize that $\nabla\simeq\nabla_{\text{e}}$ is not satisfied for all the slow solutions, because they can exist for all destabilizing $B$ values under the curve given by the continuous line, which are not necessarily $|B|\ll 1$.

Moving on to the fast solutions, we can see in Fig. \ref{Fig:SAD} that the values for $\nabla$ are significantly closer to $\nabla_{\text{ad}}$ than in the slow scenario. If $U\ll 1$ the gradients converge to $\nabla\simeq\nabla_{\text{e}}\simeq\nabla_{\text{ad}}$ to a higher level of precision. This is consistent with the analysis made in section 3 of \cite{2024ApJ...969...10C} for efficient convection ($\Gamma \gg 1$).

Summarizing, Eq. \ref{eq:velocityQ} for the dimensionless velocity $\nu$ can have either one or three real positive solutions. In particular the slow solutions have thermal gradients $\nabla$ closer to radiative gradients, whereas the fast solutions correspond to a more adiabatic regime of energy transport. This means that, for the same input parameters $\nabla_{\text{rad}},\nabla_{\text{ad}}$ and $U$ we can have coexistent solutions with not only very different values for the convective velocity but also very different regimes of energy transport.

As a final remark, we mention that whether the physical system will adopt the fast, intermediate or slow regimes likely depends on initial or boundary conditions and identifying them is beyond the scope of this work. In this work, we limit our analysis to exploring the possibility, and implications, of chemically driven convection that might develop in regions previously assumed to be stable according to the criterion given by Eq. \ref{eq:False_Ledoux_criterion}. 

\subsection{Critical lines}
\label{sub:critical}

Previously, we established the existence of fast and slow solutions to Eqs. \ref{MLT1} and \ref{MLT2}, which can coexist in some regions of the parameter space given by the gradients $\nabla_{\rm rad}$ and $\nabla_{\rm ad}$, the Ledoux term $B$, and the free-fall/thermal ratio $U$. We stated, also, the existence of two curves that delimitate the region where three solutions exist. It would be of interest, then, finding mathematical expressions for those delimiting curves. Luckily, we can do so with the help of some approximations.

Eq. \ref{eq:velocityQ}, as a quartic equation, has the following 16-term long discriminant \citep{1922jstor,math10142377}\footnote{Note that in \cite{math10142377} there is a typo in their equation A4. The correct form of the eleventh term in $\Delta_{\rm GQE}$ is $-4ac^3d^2$, not $-4ac^3d$.}
\begin{eqnarray}
\Delta_{\rm GQE} &=& 256a^3e^3 - 192a^2bde^2 - 128a^2c^2e^2 + 144a^2cd^2e  \nonumber \\
&-&27a^2d^4 + 144ab^2ce^2 - 6ab^2d^2e - 80abc^2de \nonumber \\
&+&18abcd^3 + 16ac^4e -4ac^3d^2-27b^4e^2  \\
&+&18b^3cde - 4b^3d^3 -4b^2c^3e + b^2c^2d^2, \nonumber
\end{eqnarray}

with $a=1$, $b=8U/9$, $c=(B+16U^2/9)$, $d=(B-W)8U/9$ and $e=16U^2B/9$. As in the quadratic formula, the sign of $\Delta_{\rm GQE}$ can tell us the nature of the roots, where if $\Delta_{\rm GQE}<0$ there are two distinct real solutions and two complex conjugate non real roots, and if $\Delta_{\rm GQE}>0$ the four roots are reals or none is. In the particular case of Eq. \ref{eq:velocityQ}, when there are four real solutions, one is always negative, so we consider only the three positives, since $\nu=(\nabla - \nabla_{\rm e} - B)^{1/2}>0$. The same applies when the equation has two real solutions, one is always negative, so we consider only the positive one.

Having noted the change of regime in the solutions in Fig. \ref{fig:scheme} we can try to find an expression for $\Delta_{\rm GQE}=0$ to find the two curves that delimit the number of solutions. Considering the typical values for stellar interiors, $U\ll 1$, we can neglect higher order terms in $U$ of $\Delta_{\rm GQE}$. This allow us to find an approximated expression for the first curve, which is simply given by
\begin{equation} \label{eq:uppercurve}
    W=(\nabla_{\rm rad} - \nabla_{\rm ad}) \simeq4B.
\end{equation}
This expression coincides with the limit found by \cite{1988Ap&SS.150..115U} for the maximum (positive) value for the chemical gradient $\nabla_\mu$ (or $B$ in our case) for convection to develop. In Fig. \ref{fig:scheme}, Eq. \ref{eq:uppercurve} is the upper curve that separates the region with only one convective (fast) solution from the region with three coexistent solutions.

In order to find the second critical line, we will make use of a second approximation. In Fig. \ref{fig:velocities} we can observe in the bottom panels that the change of regime from only slow solutions to three solutions occurs at $|B|\ll |W|$. Neglecting also higher order terms in $B$ in $\Delta_{\rm GQE}$ we obtain the following expression

\begin{equation}
  W=(\nabla_{\rm rad}-\nabla_{\rm ad})\simeq B-\frac{\sqrt{3}}{4U}(-B)^{3/2}.
  \label{eq:Real_Criterion}
\end{equation}    

Fig. \ref{fig:velocities} shows that Eq. \ref{eq:Real_Criterion} provides an excellent approximation for this critical line for $U \ll 1$. Eq. \ref{eq:Real_Criterion} defines an implicit relation $B^{\rm critical}=-D(W,U)$.

\subsection{Physical relevance of the existence of fast solutions}
The main consequence of  the previous result is profound. The slow mixing velocities predicted by the thermohaline process might not be the only possible mixing regime above the critical line defined by Eq. \ref{eq:Real_Criterion}. 
In those layers where the chemical gradient ($B$) lies above this critical line given by Eq. \ref{eq:Real_Criterion}, a fast convective solution is possible.
Whether this fast convective solution is established and maintained will depend on the initial conditions and perturbations, that might set the typical size of the instabilities ($l_m$),
and on whether the sources causing the inversion of the chemical gradient can keep it at a fixed value. It is worth noting that this fast mixing would, due to the low values of $U$ in stellar interiors, lead to the formation of an almost adiabatic temperature gradient, as is typically the case in deep convective zones. 

As it is clear from Fig. \ref{fig:velocities} the division of the $W-B$ space into regions where only slow termohaline-like solutions exist and regions where fast convective solutions are possible, strongly depends on the value of $U$. Moreover, this transition occurs at lower values for $|B|$ when $U$ becomes smaller. The actual value of $U$ in stellar interiors is given by  
 $U= \bar{U}(H_P/l_m)^2$. In order to use  Eq. \ref{eq:Real_Criterion} in stellar evolution computations to check for the development of convection one needs to assume the typical mixing length (i.e. size of convective elements). Given the very small values of $\bar{U}$ in stellar interiors even the assumption of convective elements much smaller than $H_P$ could result in fast adiabatic convection  being viable in a much larger region of the $W-B$ space than predicted by Eq. \ref{eq:False_Ledoux_criterion} (see next section). 

 Noteworthy, the typical size of the thermohaline instabilities, $l_{\rm therm}=(\kappa_T\,\nu/N^2)^{1/4}$, is extremely small and could lead to higher values of $U$, where only a single solution for the velocity exists,  and  the transition from thermohaline-like to convective-like velocities is smooth for the typical values of the Ledoux term $B$ and $W$ (see upper panels in Fig. \ref{fig:velocities}). Consequently, those instabilities are completely unable to drive the potential development chemically-driven adiabatic convection. As we will discuss in the next section, large-scale perturbations caused by overshooting from neighboring convective zones might in some cases create the conditions for the development of chemically-driven adiabatic convection.
This idea could offer a simple explanation for the behavior previously observed in hydrodynamical simulations.

The results presented above indicate that fast convective motions can develop in stellar interiors in the presence of much smaller chemical gradients than previously believed. Noteworthy, such situations appear to have been found in earlier studies. As noted in the introduction, hydrodynamical simulations of both the helium-core flash \citep{2011ApJ...743...55M} and the envelope mixing above the RGB bump (RGBB) \citep{2006Sci...314.1580E} show motions that are much faster than those expected from thermohaline processes. 

In the case of the RGBB, if thermohaline mixing is assumed to be responsible for the observed abundances, the required mixing efficiency must be set much higher than that expected from hydrodynamical simulations \citep{2007A&A...467L..15C, 2010ApJ...723..563D, 2011ApJ...727L...8D, 2022ApJ...941..164F}. However, recent works by \cite{2019ApJ...870L...5H} and \cite{2024ApJ...964..184F} suggest that weak magnetic fields could provide the extra mixing needed to close the discrepancy. 

In the following sections, we will study whether steady convection can be established, and thus the local mixing length theory could be applied to model both astrophysical scenarios.

\section{Mixing above the RGB bump} \label{sub:RGBB}

The  red-giant branch bump (RGBB) is the name given to an accumulation of stars at that specific luminosity in the red giant branch \citep{1985ApJ...299..674K}. It is the consequence of the decrease in the luminosity of the star caused by the passage of the H-burning shell through the composition discontinuity left by the deepest penetration of convection into the stellar envelope \citep{1967ZA.....67..420T,1968ApJ...154..581I,2023ApJ...943...45M}. Once the burning shell crosses this discontinuity the layers above the burning shell are very homogeneous and the small increase of H caused by the $^3$He($^3$He,2p)$^4$He reaction leads to a molecular weight inversion above the H-burning shell.

\begin{figure}
   \centering
    \includegraphics[width=\columnwidth]{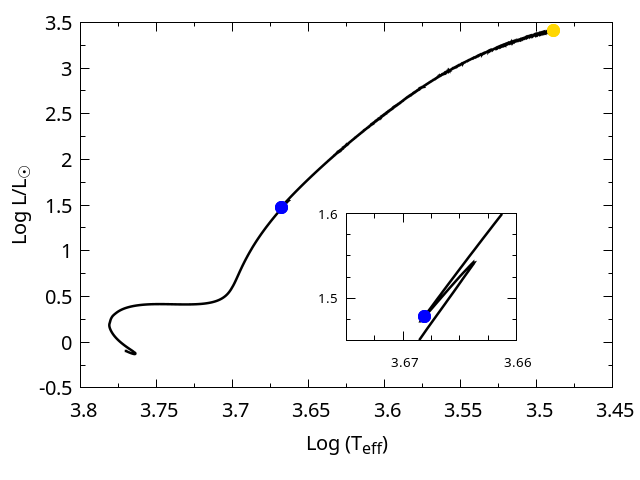}
    \caption{Hertzsrprung-Russell diagram of the evolution of a $1M_\odot$ star from the Zero Age Main sequence to the He-core flash at the tip of the RGB. The inset zooms on the location of the RGBB. Blue (yellow) point indicate the location of the snapshots of the stellar structure at the RGBB (He-core flash) discussed in this work.}
    \label{fig:HR}
\end{figure}

We have computed $1 M_\odot$ stellar models using stellar evolution code \texttt{LPCODE} \citep{2020NatAs...4...67M, 2025A&A...699A.280A}. 
{\tt  LPCODE} is a one-dimensional stellar evolution code that has been
widely used for the computation of full evolutionary sequences from
the ZAMS to the white dwarf stage. 
The last version of \texttt{LPCODE} includes
a state-of-the-art treatment of atomic, molecular and conductive
opacities as well as a detailed treatment of stellar winds and
convective boundary mixing.
The models were calculated from the evolution of models with an initial mass of $1M_\odot$ on the Zero Age
Main Sequence, through the RGBB, and up to the end of the RGB where the He core flash
develops (see Fig. \ref{fig:HR}).  The initial metallicity of the models was taken as
$Z=0.01$. Models were computed under very standard assumptions for
convection, i.e., Schwarzschild Criterion plus MLT, with no additional convective boundary mixing unless otherwise stated. Thermohaline mixing was included in the evolution as in \cite{2014A&A...570A..58W} adopting the state-of-the-art prescription of \cite{2013ApJ...768...34B}. Recent studies \citep{2019ApJ...870L...5H,2024ApJ...964..184F} indicate that such prescription might underestimate thermohaline mixing efficiency due to the action of weak magnetic fields which were not included in those previous works. We will discuss this issue at the end of this section. 

\begin{figure}[t]
   \centering
    \includegraphics[width=\linewidth]{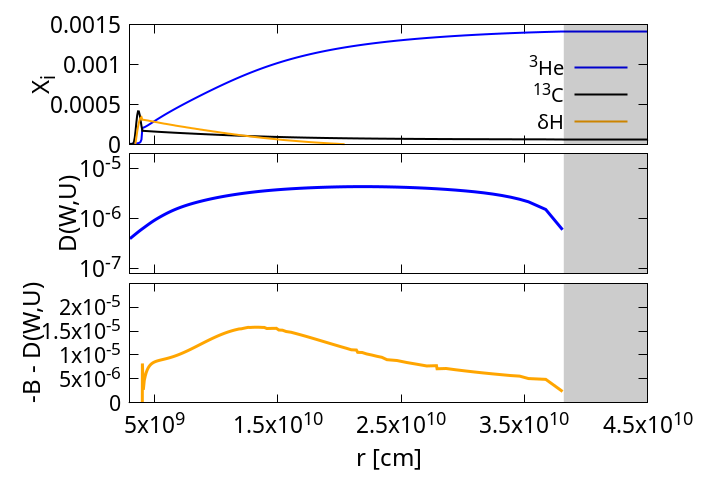}
 \caption{Physical conditions in the stellar interior as a function of the radius $r$ after the bump in the RGB. Upper panel: Abundances of H (above its surface value $\delta$H=H-H$_{\rm sur}$), $^{3}$He, and $^{13}$C. Medium panel: Layer-by-layer value of the critical value of $-B$ for the onset of (fast) chemically driven convection. Lower panel: Actual value of the instability discriminant $-B-D(W,U)$.  The inner regions that are either stable ($B>0$) or unstable against thermohaline processes ($B<0$ and $|B|<|D(W,U)|$), are not shown because they are out of scale in this plot.  The values of $D(W,U)$ have been computed with the local value of $\bar{U}$ and the assumption of $l_m= H_P$ (see text).  Grey zones indicate those regions that are already unstable to thermal convection according to the Schwarzschild criterion.}
    \label{fig:stellar_structure_RGBB}
\end{figure}

\begin{figure*}[t]
    \centering
    \includegraphics[width=\textwidth]{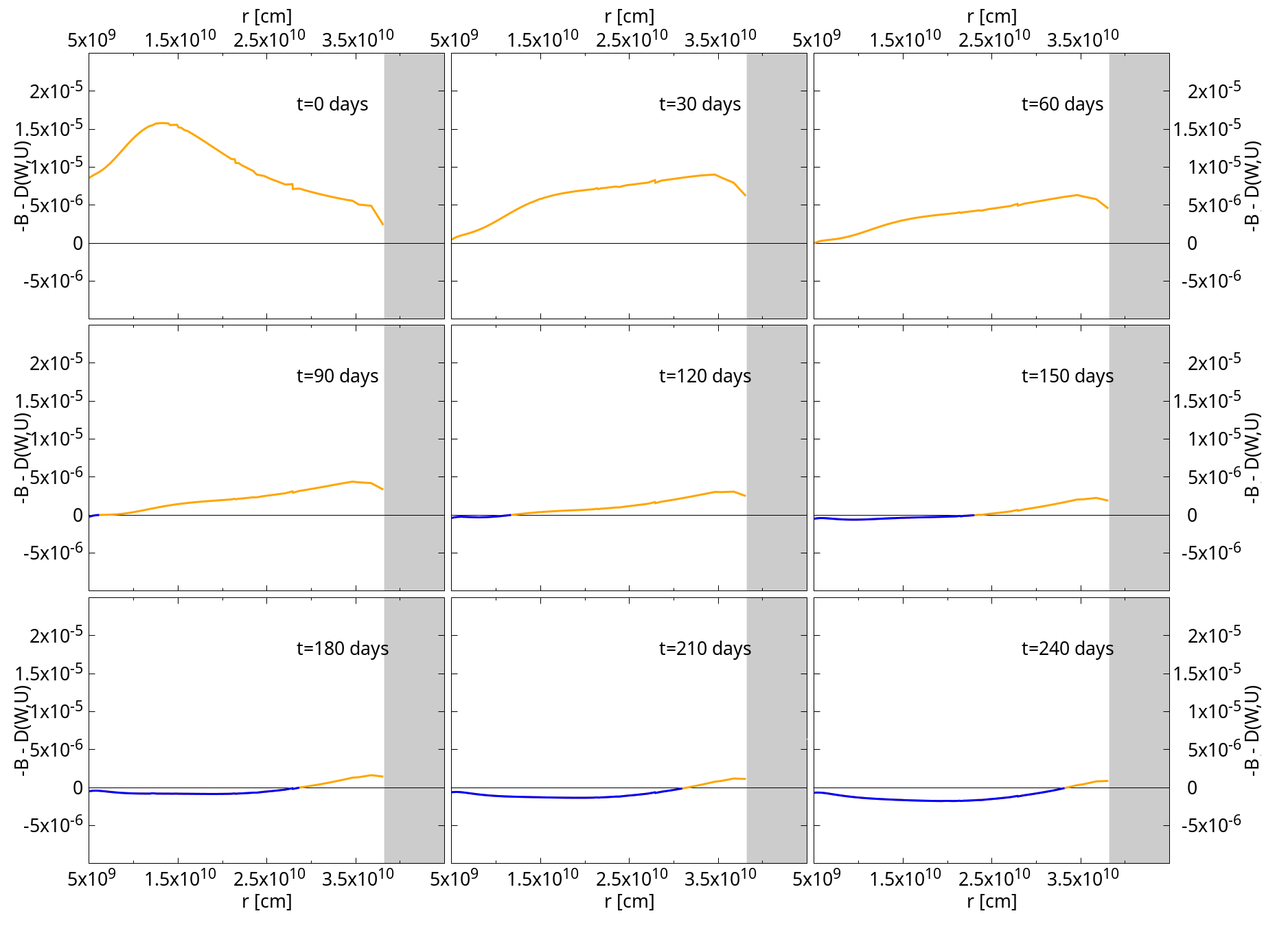}
    \caption{Temporal evolution of the chemical instabilities after the RGBB with the mixing artificially increased to fast convective-like velocities. In a span of a few months, the region is homogenized and, thus, $B$ becomes lower than the $B_{\text{crit}}$ defined by $D(W,U)$ in a short period of time. Blue and yellow are used to highlight the regions in the curve that would be stable or unstable according to Eq. \ref{eq:Real_Criterion} respectively.}
    \label{fig:RGBB_2}
\end{figure*}

The upper panel of Fig. \ref{fig:stellar_structure_RGBB} shows the chemical stratification at the bottom of the hydrogen-rich envelope at the RGBB, once the thermohaline region comes into contact with the outer convective envelope \citep[see][ for a description of the process]{2014A&A...570A..58W}. 
The relevant layers are characterized by $\bar{U}$ values in the range of $10^{-9}\hbox{--}10^{-8}$. The upper panel in Fig. \ref{fig:stellar_structure_RGBB} shows the chemical profile, the middle panel of shows the value of $D(W,U)$ in each layer, and the bottom panel shows the layer-by-layer calculation of the stability discriminant ($-B-D(W,U)$) for the case when $l_m=H_P$.

Note, however, that the typical length scales of thermohaline processes in these layers are of the order of tens for meters \citep[see also][]{2011ApJ...728L..29T}, which results in values of $U\sim 100$. Clearly, for thermohaline instabilities, motions are expected to be slow and no overlapping solutions exist.
Consequently, whether fast mixing is possible or not depends on the existence of perturbations which are large enough to create moving elements of sufficiently large size. Considering the values of the chemical gradients induced by the $^3\text{He}(^3\text{He},2\text{p})^4\text{He}$ the size of the moving elements needs to be $l_m > 10^{-1}H_P$ in order to allow fast solutions within our model.

Noteworthy, \cite{2013ApJ...769....1V}  \cite[see also][]{2002ApJ...570..825B,2023MNRAS.522.1706B} found that plumes penetrating downwards from the convective envelope have typical sizes and length scales of the order of $H_P$.

For such moving elements of size $l_m\simeq H_P$ Fig. \ref{fig:stellar_structure_RGBB} shows that the layers below the convective envelope can
 become unstable against chemically driven convection, as indicated in the lower panel of Fig.  \ref{fig:stellar_structure_RGBB}.
  In the upper panel of Fig. \ref{fig:stellar_structure_RGBB} we observe that the tiny bump in the H abundance, $\delta H = X_H - X^{\rm sup}_H \lesssim 5 \times 10^{-5}$ (where $X^{\rm sup}_H$ is the surface abundance of H) created by the $^{3}{\rm He}(^3{\rm He}, 2p)^4{\rm He}$ reaction, results in a negative value of $B$ with $|B| \lesssim 10^{-4}$ (middle panel of Fig.  \ref{fig:stellar_structure_RGBB}). Although small, this value of $|B|$ is still larger than the critical line by approximately two orders of magnitude in the case of $l_m\simeq H_P$ (see middle panel in Fig.  \ref{fig:stellar_structure_RGBB}).

Up to this point, we have established that, in a given moment in the RGBB, some layers of the star might be unstable to chemically driven convection according to Eq. \ref{eq:Real_Criterion}. If this is the case, this result would be in agreement with the fast mixing plumes observed in the hydrodynamical simulations of \cite{2006Sci...314.1580E}. However, in order to be able to describe such motions with our local mixing length theory, steady state convection needs to be established.

When a chemically unstable region is mixed, the chemical gradients diminish and the layers homogenize. In the absence of thermal instabilities, the process will be self-quenching unless there is a continuous source of the chemical instability. For example, \cite{2024ApJ...969...10C} explore a case where the compositional flux induced by the crystallization front in white dwarfs allows steady chemically driven convection to be sustained. In our case of study, the $^3\text{He}(^3\text{He},2\text{p})^4\text{He}$  reaction needs to be strong enough to sustain the instabilities for a sufficiently large number of convective turnover times.

To incorporate chemically-driven convection into a stellar evolution code, that self-consistently includes the new instability criterion and its potential impact on the thermal structure of the star, is numerically challenging and out of the scope of the present work. We can, however, easily test whether large enough chemical gradients can be established, and sustained under, the action of the strong homogenizing mixing caused by chemically driven convection.

After the RGBB was reached, we ran a simple experiment to test if steady convection can be considered (and thus MLT can be applied) or whether the extra mixing will rapidly smooth the chemical profile below the criterion given by Eq. \ref{eq:Real_Criterion}. Considering the situation shown in Fig. \ref{fig:stellar_structure_RGBB} as an initial time $t_0$, we have artificially enhanced the thermohaline efficiency below the convective envelope by several orders of magnitude to match the adjacent fast-convective velocities, where the diffusion coefficient is of the order $D\sim10^{14} \text{cm}^2/\text{s}$. With this enhanced mixing the simulation was carried over the spam of 8 months until the chemical profile was smoothed (see Fig. \ref{fig:RGBB_2}). The convective turnover time was estimated at the base of the envelope as $\tau\sim l_m/v_c$, where $l_m\simeq H_P \sim 2\times10^{10}\text{cm}$ and $v_c\sim D/l_m$, resulting in $\tau\sim 30$ days.

Taking into account this value for $\tau$, the extra mixing erases the chemical instability in less than 10 turnover times. Thus, the destabilizing H is not created fast enough in order to compensate the stabilizing effect of the mixing, and the process is self-quenching. We remind the reader that the simulation considered in this experiment did not include overshooting or enhanced mixing by magnetic fields, which are physically present in the recent  hydrodynamical simulations (which are far more sophisticated than our simple experiment).

In order to ponder the effect of convective boundary mixing (overshooting) at the bottom of the envelope on our previous result, an additional set of simulations was performed. These numerical experiments were performed with an overshooting parameter $f=0.015, \ \text{and} \  0.05$, according to the inferred values by \cite{2018ApJ...859..156K} and \cite{2025A&A...700A.163B}. Despite having an stabilizing effect over the regions considered, we find that the overall results are not significantly modified by the inclusion of overshooting, and the main conclusion remains the same. 

Moreover, considering the recent works by \cite{2019ApJ...870L...5H} and \cite{2024ApJ...964..184F}, we have made an extra set of simulations enhancing thermohaline mixing efficiency by 2 orders of magnitude, in order to roughly parametrize the extra mixing induced by weak magnetic fields. The resulting chemical gradients at the RGBB, although being flatter, are still unstable according to the criterion given by Eq. \ref{eq:Real_Criterion}. As in the case of the unmodified thermohaline mixing, this experiment shows that, if fast-convective mixing develops at this point, the chemical gradient will flatten out, suppressing the instabilities even faster than in the unmodified case.  Then, the overall conclusions are also not significantly modified.

For sake of completeness, and in light of the results discussed by \cite{2015MNRAS.446.2673L}, we performed convergence checks and found that all results presented here are fully converged.

In conclusion, although being unstable according to Eq. \ref{eq:Real_Criterion} at the moment of the RGBB, a simple steady mixing length theory is not a good framework to model this process.

\section{The He-core flash}\label{sub:HeCF}

\begin{figure*}
   \centering
   \includegraphics[width=0.49\linewidth]{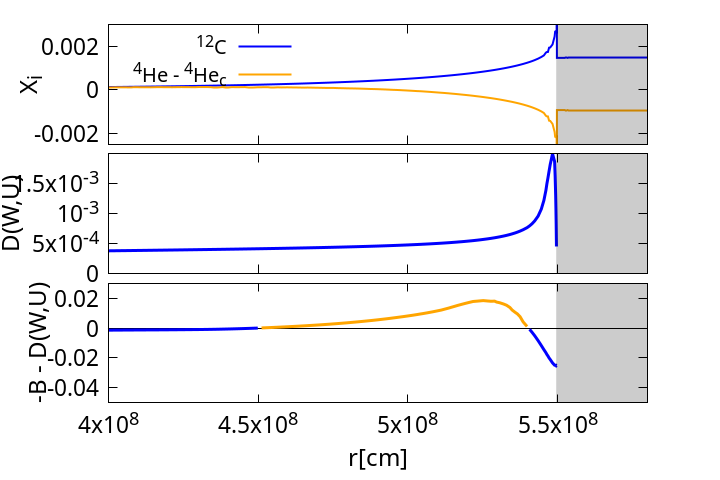}
   \includegraphics[width=0.49\linewidth]{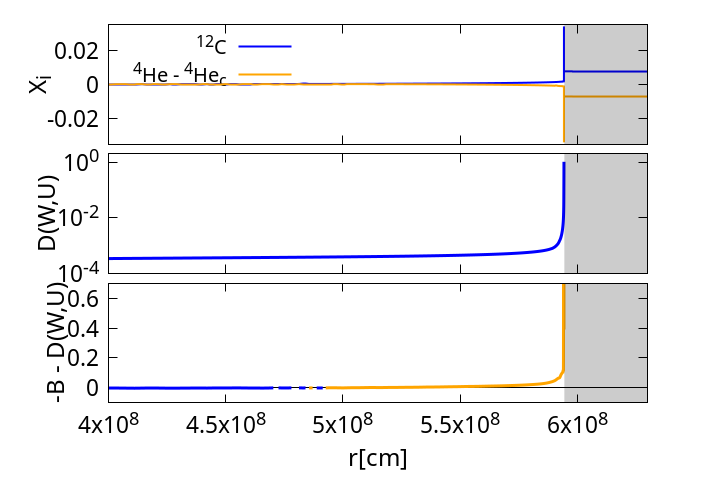}
    \caption{ Same as Fig. \ref{fig:stellar_structure_RGBB} but for the structure during the He-core flash, either at the early onset of convection (left panels) or at the moment of the maximum energy generation by the He-core flash (right panels). Lines in the upper panel indicate the abundances of $^{4}$He (relative to its value at the center $^{4}$He$_{\text{c}}$) and $^{12}$C. The values of $D(W,U)$ have been computed with the local value of $\bar{U}$ and the assumption of $l_m=10^{-4} H_P$ (see text). Yellow parts of the curve in the lower panel highlight the regions expected to develop (fast) chemically driven convective movements, while the blue ones are either stable ($B>0$) or unstable against (slow) thermohaline processes ($B<0$ and $|B|<|D(W,U)|$).}
\label{fig:stellar_structure_flash1}
\end{figure*}
\begin{figure*}[t]
   \centering
   \includegraphics[width=0.49\linewidth]{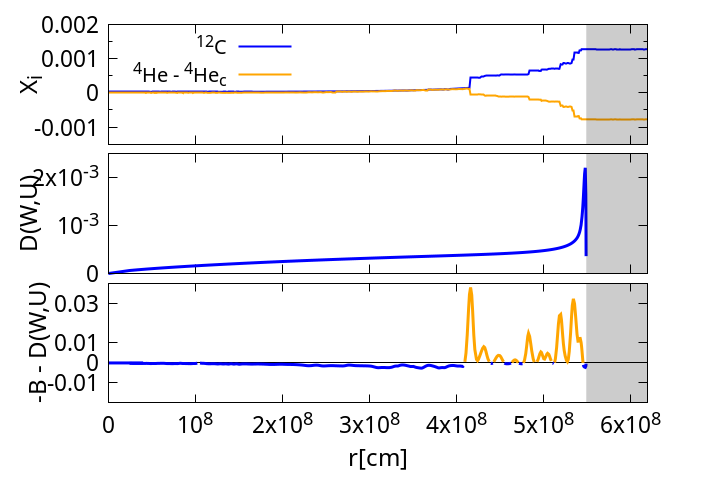}
   \includegraphics[width=0.49\linewidth]{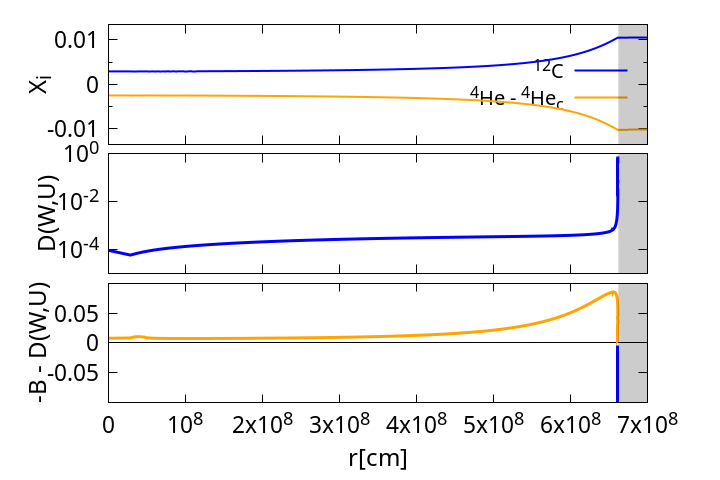}
    \caption{Same as Fig. \ref{fig:stellar_structure_flash1} but for the structure during the He-core flash when thermohaline mixing is considered in the models.}
\label{fig:structure_flash_thermo}
\end{figure*}

 Fig. \ref{fig:stellar_structure_flash1} shows the chemical profile just below the location of the flash, both at the onset of convection (left panels) and at the maximum the helium flash (right panels). As discussed in section \ref{sub:RGBB}, the models presented in Fig. \ref{fig:stellar_structure_flash1} were computed under the assumption of a bare  Schwarzschild criterion and neglecting any kind of non-convective mixing. Additionally, Fig. \ref{fig:structure_flash_thermo} shows the chemical profile when thermohaline mixing is actually included in the models. Typical values of $\bar{U}$ in these layers are 
$\bar{U} \simeq 10^{-15}\hbox{--}10^{-14}$, which imply that, even for moving elements with very small values of $l_m$, the value of $U$ would be $U \ll 1$ and fast mixing solutions could exist for chemical gradients much smaller than usually assumed when adopting eq. \ref{eq:False_Ledoux_criterion} (see Fig. \ref{fig:velocities}). Again, it should be noted as before, that typical thermohaline length scales in these layers are of the order of a few centimeters implying values of $U\sim 100$. Therefore, no fast mixing solutions exist for such instabilities within whis model. However, very simple estimation of the overshooting penetration from the nearby flash-driven convective zone shows that this process provides the required larger perturbations. Estimating the size of overshooting plumes as in \cite{2020MNRAS.493.4748W} and \cite{2017MmSAI..88..248L} we can see that overshooting perturbations are at least of the order of tens of meters ($\sim 10^{-5} H_P$), which would correspond to $U\simeq 10^{-3}$. Noteworthy this estimation is very likely a significant underestimation of overshooting \citep{2020MNRAS.493.4748W}.
The middle panels of  Figs.  \ref{fig:stellar_structure_flash1} and \ref{fig:structure_flash_thermo} show the values of $D(W,U)$ in each layer, while the bottom panels shows the layer-by-layer calculation of the stability discriminant ($-B-D(W,U)$). These results are presented for $l_m= 10^{-4} H_P$.
chemically-unstable regions appear below the convective regions at the maximum of the He-core flash even for values as small as $l_m=8.4\times 10^{-6} H_P$. For such small values of $l_m$, however, chemically-unstable regions do not appear from the onset of convection. Larger values of $l_m$ lead to the identification of even larger chemically-unstable regions. For values of $l_m\sim H_p$, closer to the values in thermal convection,  chemically-unstable regions appear even before the development of the thermal convective zone. It is clear from Figs. \ref{fig:stellar_structure_flash1} and \ref{fig:structure_flash_thermo} that the layers below the thermal He-core flash driven convective zone could be unstable to chemically driven convection. Although the value of $|B|$ in those regions is quite tiny, the critical line for the relevant values of $U$ is orders of magnitude lower, see middle panels in Figs.  \ref{fig:stellar_structure_flash1} and \ref{fig:structure_flash_thermo}. As more carbon is created during the development of the He-flash, the layers become progressively more unstable. The chemical structure and related variables at the maximum of the He-core flash is displayed in the right panels of Figs.  \ref{fig:stellar_structure_flash1} and \ref{fig:structure_flash_thermo}. It is worth noting that this is true even when thermohaline mixing is considered and is already transporting material downwards (Fig. \ref{fig:structure_flash_thermo}). In fact, as can be appreciated in the left panels of Fig. \ref{fig:structure_flash_thermo}, when thermohaline mixing is considered with the recipe of \cite{2013ApJ...768...34B} by the time the model reaches the maximum of the He-core flash, the whole region below the convective zone becomes unstable according to our criterion. The possible existence of chemically-unstable regions, where chemically driven convection can develop, underneath the thermal convective zone is  consistent with the new form of mixing discovered by Moc\'ak et al. \citep{2011ApJ...743...55M} in their hydrodynamical simulations of the He-core flash.

The consequences for the development of the He-flash could be significant. We expect chemically driven convection to develop early on, below the temperature maximum, once sufficient carbon has accumulated. The simulations presented above indicate that chemically-unstable regions do develop even when thermohaline mixing, which tends to homogenize chemical abundances, is considered. Whether steady state chemically driven convection can be established will depend on whether, even under the action of a fast mixing process, carbon production by the He-flash is able to keep a chemical gradient large enough to sustain convection.
 
 \subsection{Chemical gradients under the action of a strong mixing below the helium flash}
 
 As in section \ref{sub:RGBB} we can easily test whether large enough chemical gradients can be established, and sustained under, the action of the strong homogenizing mixing caused by chemically driven convection. To this end, we have recomputed the development of the He-core flash under the assumption that mixing extends all the way from the base of the He-core flash convective zone to the center of the star. This was done adopting three different constant values for the diffusion coefficient ($D=10^{12},10^{13},10^{14}$ cm$^2$/s). For reference, the typical values of $D=v\, l_m/3$ predicted by the MLT in the thermal convective zone during the He-core flash range from $D\sim 10^{12}$ cm$^2$/s in the earlier stages to $D\sim 10^{15}$ cm$^2$/s at the maximum of helium burning. In these simple experiments, mixing below the thermal He-flash convective zone was included from the moment the convective zone develops, regardless of whether our criterion indicates the region to be chemically-unstable.  Therefore, downward mixing in these experiments is overestimated.

\begin{figure}
   \centering
   \includegraphics[width=\linewidth]{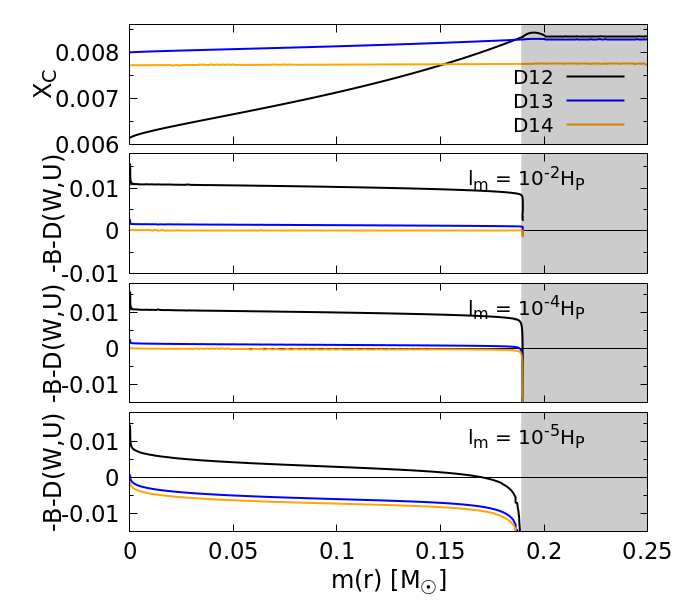}   
    \caption{$^{12}$C profiles \textit{(top panel)} and the instability discriminants $-B-D(W,U)$ \textit{(bottom panels)} tested against three different values for $l_m$ at the moment of maximum energy generation by the He-core flash. D12, D13, and D14 correspond to the experiments where mixing in the whole region below the He-flash driven thermal convective zone was included with diffusion coefficients of $D=10^{12}, \ 10^{13}$ and $10^{14}$ cm$^2$/s respectively.}
\label{fig:He-flash-D_RT}
\end{figure}

For the case in which downward mixing is assumed to happen with  $D=10^{12}$ cm$^2$/s we find that, despite this forced fast mixing, carbon production is able to keep the absolute value of the chemical gradient increasing. Assuming $l_m=H_P$ our criterion indicates that already $\sim 6200$ days before the maximum of the He-flash the whole chemical stratification below the He-core flash is  chemically-unstable, and remains so for more than ten thousand days after the maximum of the He-core flash. For smaller assumed values of $l_m$, timescales are shorter but still very significant. If we assume a value of $l_m=10^{-3} H_P$ ($l_m=10^{-4} H_P$) the whole region below the He-core flash is identified as chemically-unstable from 15 (2) days before the maximum to about 250 (16) days after the maximum of helium burning. The duration of the chemically-unstable regions is, therefore is rather long in comparison with the very fast nature of the He-flash. In particular this results suggests that, for $D=10^{12}$  cm$^2$/s unstable chemical gradients last for much longer than the typical convective turn-over time of the He-core flash (about half an hour), therefore allowing the establishment of a steady state. The chemical profile at the moment of the maximum of He burning for this simulation is shown in the upper panel of Fig. \ref{fig:He-flash-D_RT}.

For larger values of $D$, mixing happens faster and chemically-unstable regions last for a shorter period of time. For   $D=10^{13}$ cm$^2$/s, and assuming $l_m=H_P$ the whole region below the He-core flash becomes chemically-unstable from 250 days before the maximum to more than ten thousand days after the maximum. Assuming $l_m=10^{-3} H_P$ the region remains chemically-unstable from 6 days before the maximum to about 27 days after the maximum. Finally, for $D=10^{14}$ cm$^2$, and assuming $l_m=H_p$ we find that whole region below the He-flash is identified as convective from 12 days before the maximum to 1200 days after the maximum. Assuming $l_m=10^{-2} H_P$ the region remains chemically-unstable from 8 days before the maximum to about 80 days after the maximum.
It is worth noting that, for larger values of $D$ small values of $l_m$ are not physically sounding as $D=l_m v/3$ and they would imply velocities much higher than those of the thermal convective zone. The lower panels of Fig \ref{fig:He-flash-D_RT} show the chemical profiles of the simulations with $D=10^{13}$  cm$^2$/s and $D=10^{14}$ cm$^2$/s at the maximum of the He flash.

In summary these numerical experiments show that carbon production during the He-core flash is more than enough to sustain a chemical gradient that is negative enough to drive a steady state chemically driven convection, even if the lengthscale of the moving elements is orders of magnitude smaller than that of the thermal convective zone.

The main caveat of these experiments is that during these simulations we have kept the region below the helium flash with a radiative temperature gradient. However, if chemically driven convection actually develops we expect the temperature gradient in those layers to quickly become adiabatic due to the very low values of $U$ (see Fig.  \ref{fig:velocities} and Fig. \ref{Fig:SAD}). Given that in those layers $\nabla_{\rm rad}<\nabla_{\rm ad}$ this change in the temperature profile of the star would make those regions even more unstable helping the development of chemically driven convection.

It should be noted that the development of an adiabatic chemically driven convective zone, would shift the temperature maximum inward, threatening a continuous deepening of the convective zone. Such a situation could fundamentally change our understanding of the helium-core flash, potentially leading to the flash finally taking place at the center of the star and the complete disappearance of the subsequent helium sub-flashes seen in standard stellar evolution computations \citep{Kippenhahn2013, 2005essp.book.....S, 2020NatAs...4...67M}.

\section{Conclusions}
In this work, we presented a revision of the instability criteria for chemically driven convection and thermohaline mixing in the deep interior of stars. Our results show that the commonly adopted criterion (Eq. \ref{eq:False_Ledoux_criterion}) completely fails to identify the conditions under which fast chemically driven convection can develop. Furthermore, we have shown that chemically driven convection can be triggered at much smaller chemical gradients than predicted by Eq. \ref{eq:False_Ledoux_criterion} provided that the size of convective elements does not deviate by many orders of magnitude from those in regular convective zones, and that nuclear reactions sometimes are able to keep a steady chemical gradient. Moreover, we have provided a simple instability criterion (Eq. \ref{eq:Real_Criterion}) that very accurately captures the minimum required chemical gradient for fast chemically driven convection. Using this new criterion, we have revisited the conditions that arise during the helium-core flash and immediately above the RGBB. 

Our results on the RGBB indicate that regions at the bottom of the convective envelope are only marginally unstable to convection if convective elements were to be of the same size as in thermal convective regions (i.e. $l_m\sim H_P$). Moreover our numerical experiments indicate that the $^{3}{\rm He}(^3{\rm He}, 2p)^4{\rm He}$ reaction is unable to keep a large-enough molecular weight inversion once fast mixing develops. Therefore, chemically-driven convection seems to be inadequate for explaining the extra mixing at the bottom of the convective envelope of RGBB stars. A much smaller increase in mixing efficiencies, such as the enhancement of thermohaline mixing by weak magnetic fields \citep{2019ApJ...870L...5H,2024ApJ...964..184F}, seems a more promising explanation.

Conversely, our results indicate that even for very small convective elements ($l_m\gtrsim 10^{-5} H_P$) the new instability criterion identifies convectively unstable regions below the thermal He-flash driven convective zone. Moreover, our numerical experiments show that the  triple-$\alpha$ reaction is able to sustain large enough molecular weight inversions for steady-state chemically-driven convection to develop. Such chemically-driven convective zone is expected to be adiabatic. If such convective zone develops, this would have profound consequences for our understanding of the helium core-flash evolution. In particular, the new instability criterion (eq. \ref{eq:Real_Criterion}) might lead to the ignition of the He-core flash at the center, leading to the complete disappearance of the He sub-flashes seen in standard stellar evolution computations \citep{Kippenhahn2013, 2005essp.book.....S, 2011ApJ...743...55M, 2020NatAs...4...67M}.  Our theoretical findings also offer a clear physical explanation for the fast mixing processes previously observed in hydrodynamical simulations \citep{2011ApJ...743...55M}.

This revision of the instability criterion  for the development of chemically driven convection and thermohaline mixing 
could have consequences for other stages of stellar evolution, such as mixing below the flash-driven convective zones on the asymptotic giant branch \citep{2026A&A...705A..73O} or chemically-driven convection at the christallization front of white dwarfs \citep{2024ApJ...969...10C}.
This work calls for a reanalysis of the impact of inverse chemical gradients in stellar evolution computations across the entire mass range.

\begin{acknowledgments}
We thank both anonymous referees for their accurate comments and corrections that allowed us to clarify and correct previous versions of this article.
 This work was partially supported by PIP-2971 from CONICET (Argentina) and by PICT 2020-03316 from Agencia I+D+i (Argentina).
\end{acknowledgments}

\begin{contribution}

M.M.O performed the numerical computations, solving the extended MLT equations and performing the stellar simulations,
and prepared the figures.
M.M.M.B. developed the idea, identified the nature of the critical lines and derived the simplified criterion (Eq. \ref{eq:Real_Criterion}).
Both authors participated in writing the article, the design of the numerical experiments, and in the discussion of the results, their
presentations in figures and descriptions in the manuscript and in
pinpointing the conclusions.


\end{contribution}

%


\software{\texttt{LPCODE} \citep{2020NatAs...4...67M, 2025A&A...699A.280A}}



\appendix

\section{Obtaining Ledoux criterion from MLT equations}\label{methods}

With the help of Eqs. \ref{MLT1} and \ref{MLT2} it is possible to search for those solutions that are buoyantly neutral against adiabatic perturbations $\nabla-\nabla_{\rm ad}-B=0$, i.e. neutral to the Ledoux criterion. Replacing this condition into Eqs.  \ref{MLT1} and \ref{MLT2} one finds that these solutions fulfill
\begin{eqnarray}
  (\nabla_{\rm ad}-\nabla_{\rm e})^{3/2}&=& 2U (\nabla_{\rm e}-\nabla) \nonumber \\
  (\nabla_{\rm ad}-\nabla_{\rm e})^{1/2}(\nabla_{\rm e}-\nabla)&=& \frac{8U}{9} (\nabla-\nabla_{\rm rad}).
\end{eqnarray}
Ultimately, this equations can be rewritten as a single quartic equation on the variable $z=\sqrt{\nabla_{\rm ad}-\nabla_{\rm e}}$
\begin{equation} \label{eq:Ledouxquartic}
Q_2(z)=z^4+\frac{8U}{9}z^3+\frac{16 U^2}{9}z^2+\frac{16 U^2}{9}W=0.
\end{equation}  
This equation can be numerically solved to give $B^{\rm Ledoux}=D_2(W,U)$, which is shown in Fig. \ref{fig:velocities}.

To the left of the red point in Fig. \ref{fig:scheme},  the Ledoux curve falls in the 3-solutions region and only the intermediate solution fulfills $\nabla-\nabla_{\rm ad}-B=0$. Conversely, when $W$ is to the right to the red point, only one solution exist and that solution fulfills $\nabla-\nabla_{\rm ad}-B=0$.

\bibliography{NatAstron2025}{}
\bibliographystyle{aasjournalv7}



\end{document}